\documentclass[aps,prb,twocolumn,superscriptaddress,floatfix]{revtex4-2}

\usepackage{amsmath,amssymb}
\usepackage{graphicx}
\usepackage{bm}
\usepackage{xcolor}
\usepackage{hyperref}

\newcommand{\kF}{k_{\mathrm{F}}}

\newcommand{\krho}{k_{\rho}}

\newcommand{\EF}{E_{\mathrm{F}}}
\newcommand{\kTF}{k_{\mathrm{TF}}}
\newcommand{\half}{\tfrac{1}{2}}
\newcommand{\Dlt}{\Delta}
\newcommand{\xic}{\xi}

\begin{document}

\title{Analytical Charge Density Profile of Vortex Core in Weak-Coupling Superconductor}

\author{Chi-Ken Lu}
\affiliation{Department of Mathematics and Computer Science,
Rutgers University Newark, Newark, New Jersey 07102, USA}

\date{\today}

\begin{abstract}
Self-consistent Bogoliubov--de~Gennes calculations have long shown that
solving the Poisson equation inside a superconducting vortex turns a
one-signed charge depletion into a modulation that alternates in sign
with period $\pi/\kF$.  We give an elementary account of that result.
Taking the Caroli--de~Gennes--Matricon bound states in a step-like gap,
we show that the normalization of the bound-state spinor is nearly
independent of angular momentum, which collapses the mode sum into
closed form.  Inside the core the vortex winding removes one Bessel
channel from a completeness sum, so the density vanishes on the vortex
line and carries Friedel-like oscillations of wavevector $2\kF$; outside
it the sum gives a $1/r$ envelope decaying over a coherence length,
with a residual ripple.
The bound-state charge does not integrate to zero, so neutrality obliges the
extended states to compensate it exactly.  That compensation is
complete at long wavelength but fails at the diameter of the Fermi
circle, and what survives is a sign-alternating $2\kF$ modulation
reduced only by $4\kF^{2}/(4\kF^{2}+\kTF^{2})$, a factor lying between
one-half and three-quarters for any metal.  The oscillation is
therefore not a delicate effect but a consequence of neutrality and the
inefficiency of screening at large momentum transfer: in the screened
total the smooth terms cancel and only the ripple is left.
\end{abstract}

\maketitle


\section{Introduction}
\label{sec:intro}

The electronic structure of vortex cores in type-II superconductors has
been studied since Caroli, de~Gennes, and Matricon
(CdGM)~\cite{CAROLI1964307} found discrete quasiparticle bound states
at energies $E_{\mu}=\mu\Dlt^{2}/\EF$.  In conventional type-II superconductors
such as NbSe$_2$, and in the full-shell hybrid nanowires resolved
recently~\cite{PhysRevLett.134.206302}, the lowest in-gap state is
$\mu=1/2$, whereas in
topological superconductors~\cite{Sato_2017} the $\mu=0$ vortex bound
states~\cite{PhysRevB.44.9667,PhysRevB.78.132502,PhysRevB.77.220501}
are a signature of the underlying topological band
structure~\cite{PhysRevB.78.195125}. These CdGM states govern the
local spectroscopic structure of the mixed
state~\cite{doi:10.1126/sciadv.adh9163,Chen:2018aa,MAGGIOAPRILE20231354386}
and underlie a range of phenomena from spectral flow to vortex
dynamics~\cite{RevModPhys.66.1125,PhysRevLett.75.3736,Kopnin2002}.

The CdGM states arise in much the same way as the bound states of a
one-dimensional potential well.  For the well problem, the classically
allowed region lies between the two turning points at which the
classical momentum $p(x)=\sqrt{E-V(x)}$ vanishes, and the eigenenergy
$E$ follows from a quantization rule in integral
form~\cite{bender1999advanced}.  For a simple $s$-wave superconductor
in two dimensions the CdGM states are two-component spinors, and they
are likewise trapped beyond a turning point set by the centrifugal
potential and the angular momentum $\hbar\mu$.  What differs from the
potential-well problem is that the second turning point is implicit: it
occurs where the gap $\Dlt(r)$, acting in particle-hole space, meets
the energy $E$.  The seminal paper of CdGM~\cite{CAROLI1964307} and the
work of Bardeen, K\"ummel, Jacobs, and Tewordt
(BKJT)~\cite{PhysRev.187.556} determine the bound-state energy by
matching the phases accumulated by the spinor wavefunction from the
vortex line and from deep inside the bulk.

The gap profile $\Dlt(r)$ not only binds the CdGM states, it also
drives a charge
redistribution~\cite{VANDERMAREL199035,PhysRevB.46.14245}.  A simple
explanation for the charge transfer is that electrons lower their
energy by moving from the normal region inside the core into the
surrounding superconducting
subsystem~\cite{PhysRevLett.75.1384}.  The amount of transferred charge
has been estimated from the hole component of the extended states,
$n=\sum_{E_k>0}|v_k|^{2}$~\cite{VANDERMAREL199035,PhysRevB.46.14245}.
Blatter \textit{et al.}~\cite{PhysRevLett.77.566}, in the same spirit
as the charged-vortex picture of Ref.~\cite{PhysRevLett.75.1384},
estimated the local charge density $n(r)$, derived the corresponding
Poisson equation, and proposed an experiment to detect the electric
dipole generated by the vortex.

Because $\Dlt(r)$ varies slowly on the scale $\kF^{-1}$, the density
$n(r)$ obtained in Ref.~\cite{PhysRevLett.77.566} is smooth on the
scale of the coherence length $\xic$.  Any rapid variation of the
charge density must therefore come from the CdGM states.  Hayashi,
Ichioka, and Machida~\cite{doi:10.1143/JPSJ.67.3368} established
numerically that this short-wavelength inhomogeneity is encoded in the
CdGM wavefunctions and is accessible in principle by STM.  Machida and
Koyama~\cite{PhysRevLett.90.077003} performed full BdG~$+$~Poisson
calculations and found that the \emph{screened} charge changes
sign---an observation that still lacks an analytical explanation.  Friedel oscillations associated with vortex bound states
have since been resolved by STM in the iron-based superconductor
KCa$_2$Fe$_4$As$_4$F$_2$~\cite{PhysRevLett.126.257002}.  That
experiment probes the local density of states in the extreme quantum
limit $T/T_c\ll\Dlt/\EF$, and the accompanying self-consistent
calculation, performed at $\kF\xic\simeq6$, finds that the order
parameter itself acquires a Friedel-like modulation which shifts the
bound-state energies away from the $1\!:\!3\!:\!5$ ratio.  The
weak-coupling regime considered here is complementary: at
$\kF\xic\gg1$ the gap profile is essentially rigid, and the
oscillation resides in the density.
In a related setting, simulations of ultracold Fermi
gases~\cite{Riechers:2017aa} obtain the $2\kF$ Friedel
oscillation~\cite{gabovich1978screening} analytically by summing the
squared wavefunctions confined in a quantum well.  That calculation is
tractable because the infinite-well wavefunctions in one dimension are
elementary and share a common normalization constant.
For a one-dimensional well with two turning points,
Ref.~\cite{PhysRevLett.114.050401} obtained closed-form uniform
semiclassical densities---the leading corrections to Thomas--Fermi
theory---by matching Langer--Airy forms of the WKB
wavefunction~\cite{bender1999advanced} across the turning points.
The resulting expressions remain valid through the classically
forbidden region and, notably, require no explicit sum over occupied
states.
The CdGM problem
is harder on all three counts: the decay function and the particle-hole
mixing angle obey a pair of coupled nonlinear differential equations
[Eqs.~(4.17) and (4.18) of Ref.~\cite{PhysRev.187.556}], the
normalization constant is in principle mode-dependent, and
the semiclassical wavefunctions of Ref.~\cite{PhysRev.187.556}
break down inside the turning point of the centrifugal barrier.

Our aim is explanatory rather than predictive: to recover the sign
change of Ref.~\cite{PhysRevLett.90.077003} from arguments simple
enough to be checked by hand, and thereby to identify which features of
the vortex it actually depends on.  We therefore work throughout with
the most economical model that retains those features, and take from
CdGM theory whatever can be taken rather than re-deriving it.  In
particular we adopt a step-like gap profile for an isolated vortex,
$\Dlt(r)=0$ for $r<\xic$ and $\Dlt(r)=\Dlt_{\infty}$ otherwise, and
take the spectrum $E_{\mu}=\mu(\Dlt^{2}/\EF)$ as given, with $\mu$
running from $1/2$ up to $\lfloor\kF\xic\rfloor$, where the turning
point $\mu/\kF$ reaches the core edge $r=\xic$.  The step profile
simplifies the BdG equation inside the core considerably.  In the outer
region the BKJT envelope can be linearized, with the expansion
coefficients fixed by the matching conditions at $r=\xic$.  This yields
closed-form wavefunctions on both sides and makes the normalization
integral straightforward.  We find that, through a cancellation between
competing inner and outer contributions, the normalization constant is
a weak function of $\mu$ for the majority of the CdGM ladder.  This
allows the constant to be pulled out of the mode sum as an overall
prefactor, after which properties of Bessel functions deliver the
density in closed form.  The resulting analytical $n(r)$ agrees with
our numerics, most accurately away from the immediate vicinity of the
core edge.

The paper is organised as follows.  Section~\ref{sec:model} assembles
the bound-state wavefunctions---the radial equations with their exact
Bessel solutions inside the step-gap core and the linearized BKJT
solution outside, followed by the normalization constant, whose weak
dependence on $\mu$ is what makes the rest possible.
Section~\ref{sec:density} carries out the mode sum, obtaining the
in-core density from a Bessel completeness identity and the outer
profile as a decaying $1/r$ envelope.  Section~\ref{sec:electrostatics}
turns to the electrostatics: what enters Poisson's equation as source,
how the screened equation is solved by treating each wavevector in the
source separately, and what charge survives the screening, checked at
each stage against direct numerical solution.
Section~\ref{sec:discussion} sets the result beside the numerical and
experimental literature and states what is and is not claimed, and
Sec.~\ref{sec:summary} concludes.  Supplemental Material collects the
justification of the replacement trick, its error analysis, and the
verification of the solver~\cite{SM}.


\section{Model and bound-state wavefunctions}
\label{sec:model}
\subsection{Radial equations and inner solutions}

We consider a single vortex line in a weak-coupling $s$-wave
superconductor, in the notation of BKJT~\cite{PhysRev.187.556}.  After
absorbing the winding phase $e^{-i\theta}$ of the gap
$\Dlt(\mathbf{r})=\Dlt(r)e^{-i\theta}$ by the gauge rotation
$e^{-i\sigma_z\theta/2}$, the radial BdG equation reads
\begin{multline}
\sigma_z\frac{\hbar^{2}}{2m}
  \!\left[\frac{d^{2}}{dr^{2}}+\frac{1}{r}\frac{d}{dr}
         -\frac{\bigl(\mu-\frac{\sigma_z e r}{\hbar c}A_{\theta}\bigr)^{2}}{r^{2}}
         +\kF^{2}+\sigma_zE\right]\!\hat{f}\\
  =\sigma_{x}\Dlt(r)\hat{f}
  \label{eq:radial}
\end{multline}
where $\sigma_z$ and $\sigma_x$ are Pauli matrices in particle-hole
space. 
In this paper we fix the in-plane Fermi wavevector to be $\kF$.  Single-valuedness of the spinor
$\hat{f}=(f_+,f_-)^{\mathsf{T}}$ after the gauge rotation forces $\mu$
to be a half-integer ($2\mu$ odd); $f_+$ and $f_-$ are the electron and
hole amplitudes.  The azimuthal vector potential $A_\theta$ carries
both the quantized vortex flux and the applied field $H$ through
$r\int d\theta\,A_\theta=\Phi_0+\pi r^2 H$ with $\Phi_0=hc/(2e)$.
BKJT~\cite{PhysRev.187.556} split $A_\theta$ into the singular vortex
piece $\hbar c/(2er)$ and a smooth field-induced remainder
$A'_\theta$; in the dilute-vortex regime $H\ll H_{c_2}$ the latter is
negligible.

The zero-temperature charge density is the sum of squared
hole amplitudes over all positive-energy bound states,
\begin{equation}
  n(\mathbf{r})=2\sum_{E>0}|f_{E,-}(\mathbf{r})|^{2},
  \label{eq:density}
\end{equation}
the factor of $2$ accounting for spin degeneracy.  Continuum states
($E>\Dlt_\infty$) produce a spatially uniform background that is
absorbed into the bulk density $n_0$; only the localized deviation
$\delta n(r)=n(r)-n_0$ due to the CdGM bound states concerns us here.
Here $f_{E,-}$ is the radial amplitude fixed by
Eq.~\eqref{eq:norm_def} below, the angular factor
$e^{i\mu\theta}/\sqrt{2\pi}$ carrying its own normalization, so that the
areal density is Eq.~\eqref{eq:density} divided by $2\pi$.  Since the
two-dimensional bulk density is $\kF^{2}/2\pi$, densities quoted in
units of $\kF^{2}$ are then directly the modulation relative to the
bulk, and $\int\delta n\,r\,dr$ counts the electrons bound to the
vortex.

To solve Eq.~\eqref{eq:radial}, BKJT~\cite{PhysRev.187.556} introduce
the slow-mode ansatz
\begin{equation}
  \hat{f}(r) = C_{\mu}\,g(x)\,H_{\mu}^{(1)}(\kF r) + \mathrm{c.c.},
  \label{eq:bkjt_ansatz}
\end{equation}
where $H_{\mu}^{(1)}$ is the Hankel function of the first kind and
$g(x)=[e^{i\eta/2},e^{-i\eta/2}]^{\mathsf{T}}e^{-\xi_{1}}$ is a spinor
envelope carrying the electron-hole mixing angle $\eta$ and the
amplitude-decay function $\xi_1$.  The normalization constant
$C_{\mu}$ is fixed by
\begin{equation}
  C_{\mu}^{2}\int_{0}^{\infty}r\,dr\,
  \bigl(|f_{+}|^{2}+|f_{-}|^{2}\bigr)=1.
  \label{eq:norm_def}
\end{equation}
Only the mode index need be carried: the Fermi surface is a single
circle and no $k_z$ integration is involved, so $\kF$ is common to
every mode.  What does enter parametrically is $\kF\xic$, through the
core size and the number of modes it admits.  A central result
established below is that at fixed $\kF\xic$ the constant depends only
weakly on $\mu$, so that it may be replaced by its mean
$\langle C^{2}\rangle$ over the ladder, which itself scales as
$(\kF\xic)^{-1}$.

The slow-mode coordinate
\begin{equation}
  x(r)=\frac{\Dlt_{\infty}}{\EF}\kF\sqrt{r^{2}-r_{t}^{2}},
  \label{eq:bkjt_x}
\end{equation}
is built from the in-plane kinetic energy
$\EF=\hbar^2\kF^{2}/2m$.  The centrifugal barrier places a
classical turning point at $r_t=\mu/\kF$, which grows linearly with
$\mu$ and separates the classically accessible region ($r>r_t$) from
the forbidden one ($r<r_t$).  Far from the core the gap saturates to
$\Dlt_\infty$, fixing the coherence length
$\xic=2\EF/(\pi\kF\Dlt_\infty)$.

The envelope functions $\eta(x)$ and $\xi_{1}(x)$ obey the coupled
first-order system [Eqs.~(4.17)--(4.18) of BKJT~\cite{PhysRev.187.556}]
\begin{align}
  \frac{d\eta}{dx}+\delta(x)\cos\eta &= \Lambda + F(x),
  \label{eq:eta_ode}\\
  \frac{2\,d\xi_{1}}{dx} &= \delta(x)\sin\eta,
  \label{eq:xi1_ode}
\end{align}
with normalized gap $\delta(x)=\Dlt(r)/\Dlt_\infty$, normalized
eigenenergy $\Lambda=E/\Dlt_{\infty}$, and vortex-flux term
$F(x)=b/(x^{2}+b^{2})$, $b=\mu\Dlt_\infty/\EF$ (the smooth
$A'_\theta$ correction is negligible inside the core).

For the step-gap model ($\delta=0$ for $r<\xic$, $\delta=1$ for
$r\geq\xic$) the nonlinear coupling in
Eqs.~\eqref{eq:eta_ode}--\eqref{eq:xi1_ode} switches off inside the
core, and the interior solution is exact,
$\eta(x)=\Lambda x+\arctan(x/b)$ with $\xi_1=0$ for
$0\leq x\leq x_c\equiv x(\xic)$.  BKJT~\cite{PhysRev.187.556} obtain
the bound-state energy $E(\kF,\mu)$ by integrating the ODEs outward
from the turning point $x=0$ and inward from the London penetration
depth $x=x_\lambda$, and matching at the core edge $x=x_c$.

The BKJT ansatz is nevertheless unsuitable for the inner region, on two
counts.  First, $x(r)$ turns imaginary for $r<r_t(\mu)$, so
Eq.~\eqref{eq:bkjt_ansatz} is ill-defined throughout the classically
forbidden region.  Second, the carrier $H_\mu^{(1)}$ cannot reproduce
the power-law behavior at the origin dictated by the vortex topology.
Both deficiencies worsen as $\mu$ grows, since the forbidden region
$r<r_t=\mu/\kF$ then covers a larger fraction of the core.

\begin{figure*}[t]
\centering
\includegraphics[width=\textwidth]{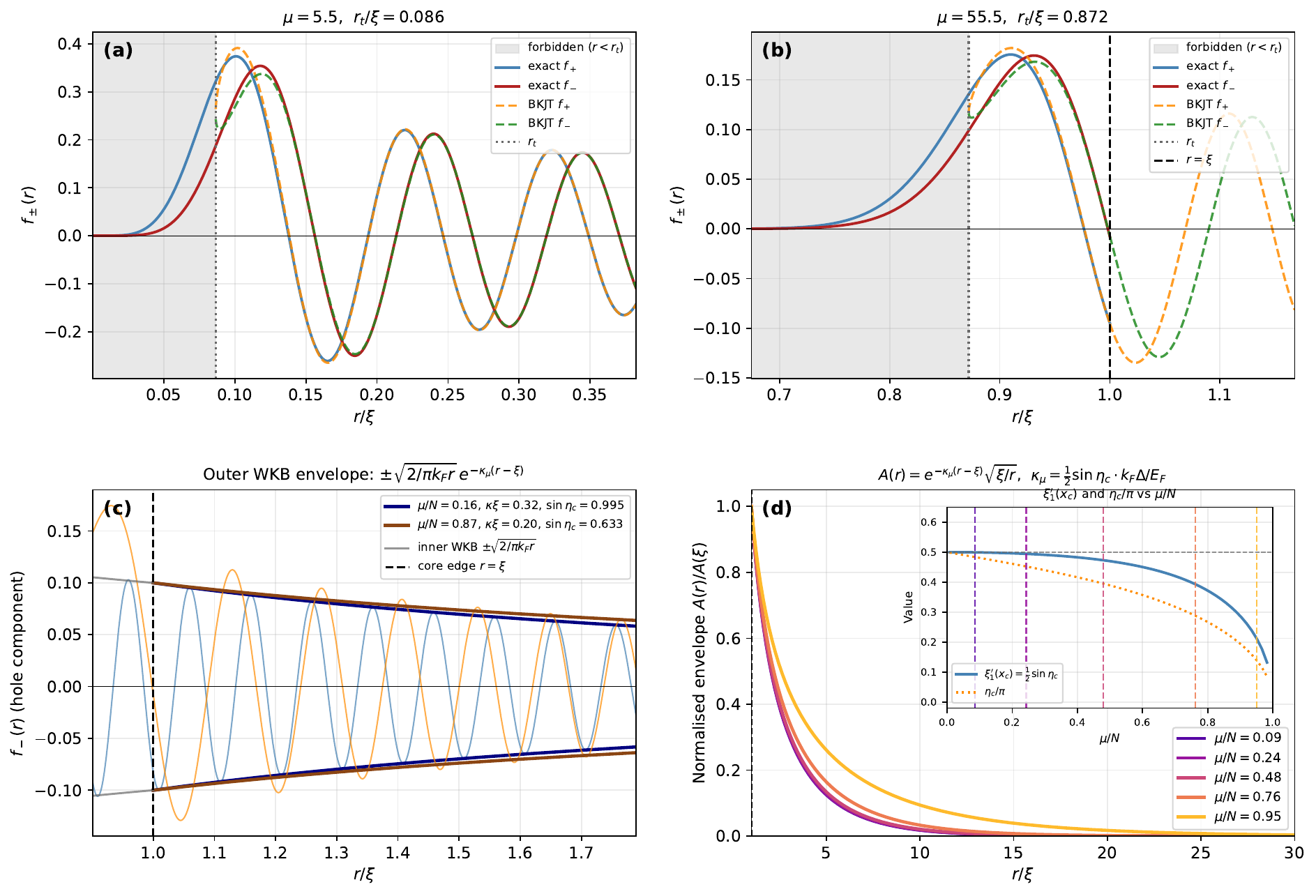}
\caption{%
CdGM bound-state wavefunctions computed with $\kF=1$, $m=\half$,
$\Dlt/\EF=0.01$, giving $\xic\approx 63.7\,\kF^{-1}$ and
$N\equiv\kF\xic\approx 63.7$ modes.
\textbf{(a)}~Electron ($f_+$, solid blue) and hole ($f_-$, solid red)
components alongside the BKJT inner approximation (dashed) for
$\mu=5.5$ ($r_t/\xic=0.086$).
The classically forbidden region $r<r_t$ is shaded grey; the turning
point (dotted vertical) lies deep inside the core.
The exact Bessel and BKJT solutions overlap closely over most of the
classically allowed region~\cite{note_turningpoint}.
\textbf{(b)}~Same for $\mu=55.5$ ($r_t/\xic=0.867$).
The viewing window crosses the core edge $r=\xic$ (vertical dashed
line); for $r>\xic$ the Taylor-linearized outer BKJT solution is
appended (dashed curves).
The large forbidden region leaves only a narrow classically allowed
inner interval, and the electron and hole envelopes separate
immediately past $r=\xic$ as the gap turns on.
\textbf{(c)}~Hole component $f_-$ in the outer region $r>\xic$ for
two representative modes (faint curves), with the WKB amplitude
envelopes shown prominently (thick curves).
The inner WKB envelope $\pm\sqrt{2/\pi\kF r}$ (grey) is displayed
for one period inside the core; the outer envelopes
$\pm\sqrt{2/\pi\kF r}\,e^{-\kappa_\mu(r-\xic)}$ (colored) separate
immediately at the core edge because the decay rate
$\kappa_\mu=\tfrac{1}{2}\sin\eta_c\cdot\kF\Dlt/\EF$ is larger for
the low-lying mode ($\mu/N=0.16$, $\sin\eta_c\approx 1$, navy
envelope) than for the high-lying mode ($\mu/N=0.87$,
$\sin\eta_c\approx 0.71$, brown envelope).
\textbf{(d)}~Normalized outer envelopes
$A(r)=e^{-\kappa_\mu(r-\xic)}\sqrt{\xic/r}$
for five modes spanning $\mu/N=0.09$--$0.95$, plotted out to
$30\xic\approx\lambda_L$.
Low-$\mu$ modes are confined within a few coherence lengths of the
core, while the highest mode barely decays over the entire London
length, illustrating the order-of-magnitude variation in $\kappa_\mu$
across the CdGM ladder.
\textit{Inset}: outer-decay coefficient
$\xi_1'(x_c)=\tfrac{1}{2}\sin\eta_c$ (solid blue) and normalized
phase $\eta_c/\pi$ (dotted orange) versus $\mu/N$.
The horizontal dashed line marks the exact low-$\mu$ limit
$\xi_1'=\tfrac{1}{2}$, corresponding to $\kappa_\mu\xic=1/\pi$; vertical dashed lines indicate the five
modes shown in the main panel, demonstrating that $\kappa_\mu\xic$
stays near $1/\pi$ for the majority of the ladder.}
\label{fig:wavefunctions}
\end{figure*}

Inside the core the gap vanishes and the BdG equations decouple.  With
wavevectors $k_{\pm}=\kF\pm q$ and $q=\kF E/(2\EF)$, the regular
solutions are exact Bessel functions
\begin{equation}
  f_{\pm}(r)=C_{\mu}\,J_{\mu\mp 1/2}(k_{\pm}r),
  \label{eq:inner_bessel}
\end{equation}
of orders $\mu\mp\half$, smooth everywhere---including across the
turning point---with no approximation required.  The inner and outer
regions are then carried by different functions, which need not join
continuously at $r=\xic$; the resulting step in the density is a few
percent and is confined to a shell about the core edge, from which no
quantitative result below is drawn~\cite{note_coreedge}.
Figures~\ref{fig:wavefunctions}(a) and~(b) benchmark these against the
BKJT approximation for a low-lying mode ($\mu=5.5$, $r_t/\xic=0.086$)
and a high-lying mode ($\mu=55.5$, $r_t/\xic=0.872$).  Outside the core
($r>\xic$, $\delta=1$) the envelope is linearized in $\delta x=x-x_c$,
\begin{align}
  \eta(x) &\approx \eta_{c}+\eta'(x_c)\,\delta x,
  \label{eq:eta_taylor}\\
  \xi_1(x) &\approx \xi_1'(x_c)\,\delta x,
  \label{eq:xi1_taylor}
\end{align}
subject to $\xi_1(x_c)=0$.  The coefficients at
$x_c=\sqrt{(4/\pi^2)-b^2}$ follow from
Eqs.~\eqref{eq:eta_ode}--\eqref{eq:xi1_ode}: for low-lying states
($E\ll\Dlt_{\infty}$) one finds
\begin{eqnarray}
  \eta_c&\approx&\pi/2-b(\pi^2-4)/(2\pi)\:,\\
  \xi_1'(x_c)&\approx&\frac{1}{2}\cos\bigl(b(\pi^2-4)/(2\pi)\bigr)\:.
\end{eqnarray}
Truncating $\xi_1$ at first order is justified by the bound
$|d\xi_1/dx|\leq\frac{1}{2}$ implied by Eq.~\eqref{eq:xi1_ode}:
retaining higher-order terms would violate this constraint.

The mode structure follows a clear pattern shown in the top two panels of Fig.~\ref{fig:wavefunctions}. 
The components $f_{E,\pm}$
rise from the origin as $r^{\mu\mp 1/2}$, oscillate with period
$2\pi/\kF$ in the classically allowed window $r_t<r<\xic$, and then
decay exponentially beyond the core edge.
The fact that $f_-(0)=J_{\nu_-}(0)=0$ for every mode
($\nu_-=\mu+\half\geq 1$) is a direct consequence of the vortex
winding while the electron component reaches the origin only for
$\mu=\half$, where $f_+(0)=J_0(0)=1$.
The deacying into the bulk beyond the core edge is controlled by,
\begin{equation}
  \kappa_\mu\equiv\xi_1'(x_c)(\kF\Dlt_{\infty}/2\EF)=
  \frac{\sin\eta_c}{\pi\xi}
  \,\label{eq:kappa_xi}
\end{equation}
From the panel (c), the CdGM modes
decay on the scale $\pi\xic$ and the weakly $\mu$-dependent function $\sin\eta_c$. 
The inset of Fig.~\ref{fig:wavefunctions}(d) plots
$\xi_1'(x_c)=\tfrac{1}{2}\sin\eta_c$ and $\eta_c/\pi$ against $\mu/N$,
confirming that $\kappa_\mu\xic$ stays near $1/\pi$ across most of the
ladder and departs appreciably only for $\mu/N\gtrsim 0.7$. 
The spread in
outer decay rates is illustrated in Figs.~\ref{fig:wavefunctions}(c)
and~(d): panel~(c) shows the hole-component WKB envelope
$\pm\sqrt{2/\pi\kF r}\,e^{-\kappa_\mu(r-\xic)}$ for two representative
modes just past the core edge, while panel~(d) extends the comparison
to five modes across $\xic\leq r\leq 30\xic\approx\lambda_L$.
Those characters of the CdGM modes
are crucial for the discussion of normalization constant.


\subsection{Normalization of the modes}

As $\mu$ increases along the ladder, the classically allowed window
$\xic-r_t$ narrows, which reduces the inner contribution to the norm,
while $\kappa_\mu$ decreases as $\sin\eta_c\to 0$, which increases the
outer contribution.  These two trends largely cancel.  Approximating
the oscillating integrands by their asymptotic cycle averages,
$\langle J_\nu^2(\kF r)\rangle\to 1/(\pi\kF r)$ inside and
$\langle f_+^2+f_-^2\rangle\to 2e^{-2\kappa_\mu(r-\xic)}/(\pi\kF r)$
outside, the normalization integral Eq.~\eqref{eq:norm_def} splits into
\begin{align}
  I_\mathrm{in}(\mu) &\approx \frac{2(\xic-r_t)}{\pi \kF}
    = \frac{2\xic}{\pi \kF}\!\left(1-\frac{\mu}{N}\right),
    \label{eq:Iin}\\[3pt]
  I_\mathrm{out}(\mu) &\approx \frac{1}{\pi \kF\kappa_\mu}
    = \frac{2\EF}{\pi \kF^{2}\Dlt_{\infty}\sin\eta_c}.
    \label{eq:Iout}
\end{align}

\begin{figure*}[t]
\centering
\includegraphics[width=\textwidth]{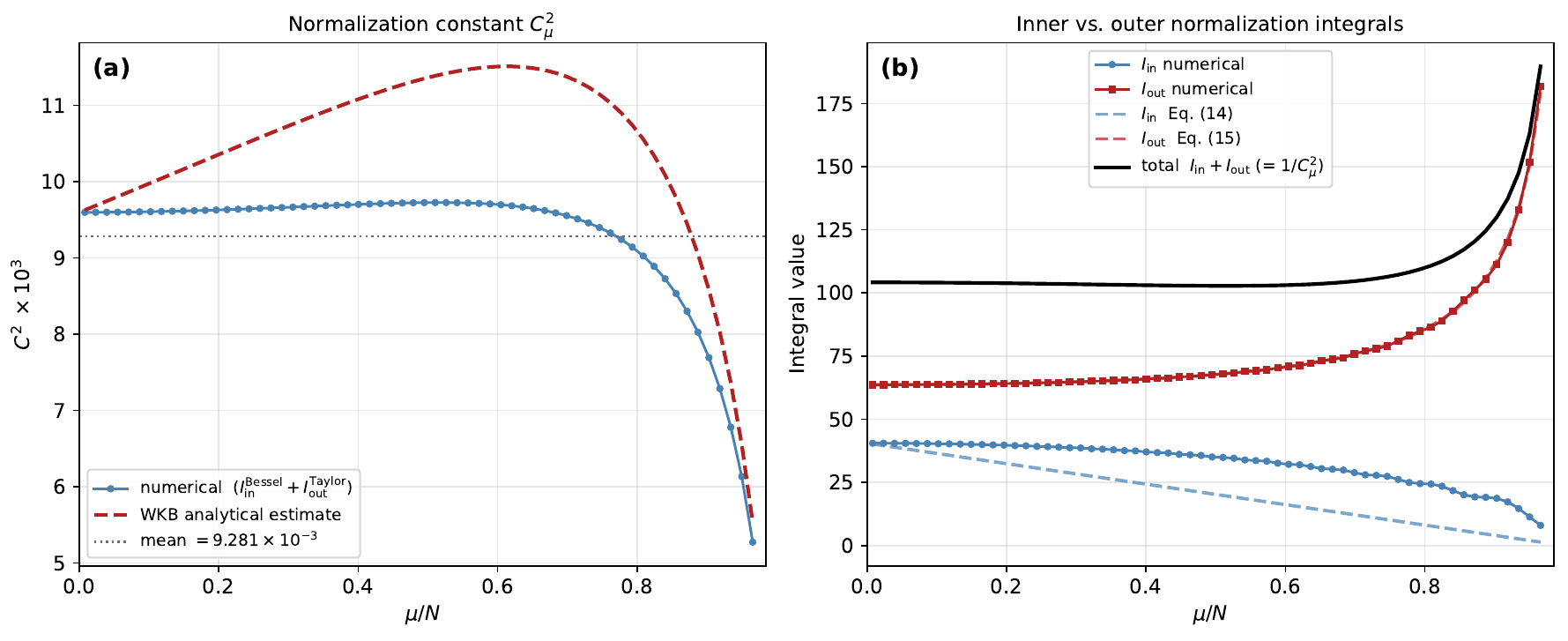}
\caption{%
Normalization constant $C_{\mu}^{2}$ and its decomposition into inner and
outer partial integrals, computed with $\kF=1$, $m=\half$,
$\Dlt/\EF=0.01$ ($N\approx 63.7$ modes).
\textbf{(a)}~$C_{\mu}^{2}$ versus $\mu/N$.
The solid blue curve (numerical: exact Bessel inner $+$
Taylor-linearized BKJT outer, integrated to $30\xic$) remains within
$6\%$ of the mean $\langle C^2\rangle\approx 9.3\times 10^{-3}$
(dotted line) for $\mu/N\lesssim 0.75$, confirming the weak dependence
on $\mu$.
The dashed red curve is the WKB analytical estimate
$(I_\mathrm{in}+I_\mathrm{out})^{-1}$ from
Eqs.~\eqref{eq:Iin}--\eqref{eq:Iout}, which captures the qualitative
trend but overestimates $C_{\mu}^{2}$ at intermediate $\mu$ because the WKB
cycle average underestimates $I_\mathrm{in}$ by neglecting the Bessel
function's exponentially small tail in the classically forbidden
region $r<r_t$.
Both curves drop sharply for $\mu/N\gtrsim 0.8$ as $\kappa_\mu\to 0$
and the outer wavefunction ceases to be normalizable within $30\xic$.
\textbf{(b)}~Partial integrals $I_\mathrm{in}$ (blue, decreasing) and
$I_\mathrm{out}$ (red, increasing) versus $\mu/N$, alongside the WKB
analytical estimates (dashed,
Eqs.~\eqref{eq:Iin}--\eqref{eq:Iout}).
The solid black curve is the total $I_\mathrm{in}+I_\mathrm{out}=1/C_{\mu}^{2}$,
which is nearly flat for $\mu/N\lesssim 0.75$, making the cancellation
mechanism directly visible.
The analytical $I_\mathrm{in}$ lies below the numerical values
(dashed vs.\ solid blue) because the WKB average ignores the
non-negligible contribution of the exponentially decaying Bessel tail
in the forbidden region.}
\label{fig:C2}
\end{figure*}

Fig.~\ref{fig:C2} displays the numerical results for computing the normalization constant 
$C^2_\mu$ along with the analytical estimates. Since $I_\mathrm{out}>I_\mathrm{in}$ for all modes
[Fig.~\ref{fig:C2}(b)], the outer integral dominates $1/C_{\mu}^{2}$, and its
near-constancy for $\mu/N\lesssim 0.7$ 
is the chief reason that $C_{\mu}^{2}=1/(I_\mathrm{in}+I_\mathrm{out})$ stays within
$6\%$ of its mean $\langle C^2\rangle\approx 9.3\times 10^{-3}$ for
$\mu/N\lesssim 0.75$.  Only near the top of the ladder, where
$\kappa_\mu\to 0$ and the linearized outer solution breaks down, does
$C_{\mu}^{2}$ deviate noticeably, as Fig.~\ref{fig:C2} confirms.  


\section{Charge density profile}
\label{sec:density}
\subsection{Mode sum inside the core}

With $C_{\mu}^{2}$ nearly constant and $k_{-}\approx\kF$ (valid to leading
order in $\Dlt_\infty/\EF$), the in-core density sum
Eq.~\eqref{eq:density} reduces to
\begin{equation}
  \frac{\delta n(r<\xic)}{\langle C^{2}\rangle} =
  2\sum_{\mu=1/2}^{N-\half}J_{\mu+\half}^{2}(\kF r).
  \label{eq:density_sum}
\end{equation}
The topological constraint $\nu_{-}=\mu+\half\geq 1$ excludes the
$\nu_-=0$ Bessel channel and enforces $n(0)=0$ exactly.  For modes with
$\mu\geq N+\half$ the classically forbidden region covers the entire
core, so their contribution at any fixed $r<\xic$ is negligible and the
sum in Eq.~\eqref{eq:density_sum} may be extended to infinity.


\begin{figure*}[tb]
\centering
\includegraphics[width=\textwidth]{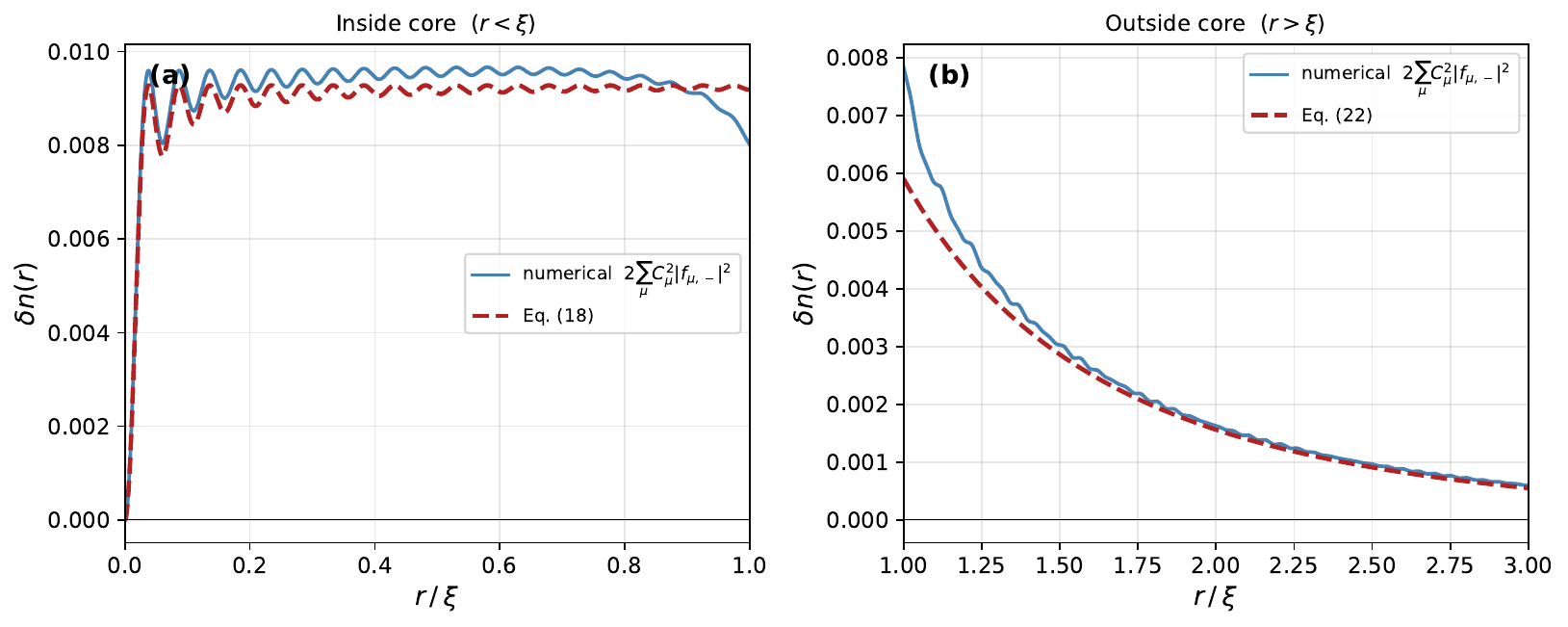}
\caption{%
Charge-density perturbation $\delta n(r)$ computed with $\kF=1$,
$m=\half$, $\Dlt/\EF=0.01$ ($\xic\approx 63.7\,\kF^{-1}$,
$N\approx 63$ modes).
Solid blue: numerical sum $2\sum_{\mu}C_{\mu}^{2}|f_{\mu,-}(r)|^{2}$
over all CdGM hole amplitudes; dashed red: analytical closed form.
\textbf{(a)}~Inner region $r<\xic$: analytical approximation
$\langle C^{2}\rangle[1-J_{0}^{2}(\kF r)]$ [Eq.~\eqref{eq:main}].
The two curves agree closely; the residual discrepancy reflects the
$\approx 6\%$ mode-to-mode variation of $C_{\mu}^{2}$ about its mean
$\langle C^{2}\rangle\approx 9.3\times 10^{-3}$ [Fig.~\ref{fig:C2}(a)].
\textbf{(b)}~Outer region $r>\xic$: the factored envelope
$2\langle C^{2}\rangle\xic/(\pi r)\,e^{-2(r-\xic)/\pi\xic}$
[Eq.~\eqref{eq:outer}].
The closed form captures the smooth $1/r$ envelope.  Two distinct
effects meet at $r=\xic$ and should not be conflated: the offset
between the dashed and solid curves there is the $\sim36\%$ WKB
artifact of the closed form at the classical turning point, discussed
in the text, whereas the small step in the numerical curve itself
($2.6\%$ here) is the discontinuity between the exact-Bessel inner and
BKJT outer constructions~\cite{note_coreedge}.
In both panels the bound-state density is positive definite;
sign-alternating oscillations appear only in the screened charge, after
the Poisson equation is solved.}
\label{fig:density}
\end{figure*}

Equation~\eqref{eq:density_sum} is a partial Bessel--Parseval sum.
Applying the completeness identity
\begin{equation}
  J_{0}^{2}(x) + 2\sum_{\nu=1}^{\infty}J_{\nu}^{2}(x) = 1,
  \label{eq:parseval}
\end{equation}
which follows from Parseval's theorem applied to the generating
function $e^{ix\sin\theta}=\sum_{\nu}J_{\nu}(x)e^{i\nu\theta}$~\cite{DLMF},
the topologically missing $\nu_-=0$ term immediately yields the central
result for $r<\xic$:
\begin{equation}
  \boxed{
    \frac{\delta n(r<\xic)}{\langle C^{2}\rangle} \approx 1 - J_{0}^{2}(\kF r).
    }
  \label{eq:main}
\end{equation}
This satisfies $n(0)=0$ by construction and evaluates to
$\delta n(\xic^-)/\langle C^{2}\rangle=1-J_0^2(2\EF/\pi\Dlt_{\infty})\approx 1$
at the core edge in the weak-coupling limit.  The point worth
emphasizing is that two physically motivated simplifications---constant
$C_{\mu}^{2}$ and $k_-\approx\kF$---are exactly what is needed to bring an
exact mathematical identity into play, collapsing an otherwise
intractable finite Bessel sum into a single closed form.
Figure~\ref{fig:density}(a) tests Eq.~\eqref{eq:main} against the full
numerical CdGM sum; the two curves agree closely throughout the core.
The density is positive definite everywhere, being a sum of squared
amplitudes with positive weights $C_{\mu}^{2}>0$; sign reversal appears
only in the \emph{screened} charge, after the Poisson equation is
solved.

Two limiting forms of Eq.~\eqref{eq:main} are instructive.  Near the
vortex axis ($r\ll1/\kF$), expanding $J_{0}(x)=1-x^{2}/4+\cdots$
gives the quadratic onset $\delta n/\langle C^{2}\rangle\approx(\kF r)^{2}/4$, which
is invisible to any coarse-grained probe.  Over most of the core
interior ($r\lesssim 0.85\,\xic$), the large-argument asymptotics
$J_{0}(x)\sim\sqrt{2/(\pi x)}\cos(x-\pi/4)$ give oscillations at
wavevector $2\kF$,
\begin{equation}
  \frac{\delta n(r)}{\langle C^{2}\rangle}\approx 1
  -\frac{1}{\pi\kF r}-\frac{\sin(2\kF r)}{\pi\kF r},
  \label{eq:friedel}
\end{equation}
in which the smooth background and the oscillation carry equal weight.


\subsection{Outside the core}

For $r>\xic$ the hole amplitude is given by the Taylor-linearized BKJT
outer solution, Eqs.~\eqref{eq:eta_taylor}--\eqref{eq:xi1_taylor},
\begin{equation}
  f_{\mu,-}(r)
  \propto e^{-\kappa_{\mu}r}
  \bigl[\cos(\eta_\mu/2)\,J_\mu(\kF r)+\sin(\eta_\mu/2)\,Y_\mu(\kF r)\bigr],
  \label{eq:outer_hole}
\end{equation}
with $\kappa_\mu$ from Eq.~\eqref{eq:kappa_xi}.  Squaring and applying
the large-argument Bessel forms leaves a common exponential envelope
multiplying a smooth term $\propto1/\kF r$ and a term oscillating at
$2\kF$.

Because $\mu$ is half-integer, adjacent modes enter the oscillating
term with opposite signs and the mode sum largely cancels it.  The
cancellation is substantial but incomplete, since $\kappa_\mu$ and the
mixing angle both drift along the ladder: a $2\kF$ ripple survives past
the core edge and decays over a few coherence lengths.  What follows is
therefore the smooth envelope of the outer density, with the ripple
riding on it, as Fig.~\ref{fig:density}(b) shows.

Replacing $C_{\mu}^{2}$ by its mode average leaves a sum over envelopes
alone,
\begin{equation}
  \delta n(r>\xic)\simeq\frac{2\langle C^{2}\rangle}{\pi \kF r}
  \sum_{\mu}e^{-2\kappa_{\mu}(r-\xic)} \:.
  \label{eq:outer_envsum}
\end{equation}
From above expression, it can be seen that the mode index $\mu$ only enters
$\sin\eta_c$ in the decaying factor $\kappa_\mu$ in Eq.(\ref{eq:kappa_xi}). 
Over the bulk of the
ladder it stays close to unity: $\kappa_\mu\pi\xic$ runs from $1$ at
$\mu=\half$ down to $0.26$ at the top, so all but the last few modes decay
at nearly the same rate.  The sum therefore factorises, leaving a
common exponential,
\begin{equation}
  \delta n(r>\xic)\;\simeq\;\frac{2\langle C^{2}\rangle\xic}{\pi r}\,
  e^{-2(r-\xic)/\pi\xic} ,
  \label{eq:outer}
\end{equation}
a $1/r$ envelope falling on the scale $\pi\xic/2$.  This is the whole
of what the outer region contributes qualitatively: the bound-state
charge is not confined to $r<\xic$ but leaks out over roughly a
coherence length.


At the edge Eq.~\eqref{eq:outer} returns $2\langle C^{2}\rangle/\pi$,
about a third below the inner boundary value
$\langle C^{2}\rangle$.  The shortfall belongs to the closed form
rather than to the mode sum: the large-argument expansion discards the
turning-point enhancement of $J_\mu$ and $Y_\mu$, which is largest for
precisely the high-$\mu$ modes that dominate at $r\simeq\xic$.  It is
unrelated to, and much larger than, the step carried by the numerical
sum itself~\cite{note_coreedge}.

In passing, the amount of charge carried by the vortex is computed
quite differently from the consideration of mass merons/Skyrmions in two-dimensional Dirac fermion systems 
in which the topological objects are formed by the elementary vortex carrying $1/2$ net electron and the total
charge is determined by the representation of the involved order parameters~\cite{PhysRevLett.108.266402}.

\section{Electrostatics of the core charge}
\label{sec:electrostatics}
\subsection{The source term}

Far from the core the electron density is that of a uniform
superconductor,
\begin{equation}
  n=2N_{\mu}\Bigl[\EF+\frac{\Dlt^{2}}{4\EF}\Bigr],
  \label{eq:n_uniform}
\end{equation}
with $2N_{\mu}$ the density of states~\cite{PhysRevB.46.14245}.  The
ionic background is rigid and cancels this exactly, so the vortex
problem is one of deviations.  Writing $\delta n_{\rm total}$ for the
departure from Eq.~\eqref{eq:n_uniform}, the ions never appear again,
and $\delta n_{\rm total}$ is the charge a local probe would measure.
It is the source of Poisson's equation in its bare form,
\begin{equation}
  \nabla^{2}\phi=-\frac{4\pi e}{\varepsilon}\,\delta n_{\rm total}.
  \label{eq:poisson_bare}
\end{equation}

Equation~\eqref{eq:n_uniform} already fixes one contribution.  Inside the vortex core
the gap is suppressed so the density falls by
$2N_{\mu}\Dlt^{2}/4\EF=\tfrac{1}{4}n(\Dlt/\EF)^{2}$, 
leaving the core positively charged
as the charged-vortex picture
requires~\cite{PhysRevLett.75.1384}.  
Since Eq.~\eqref{eq:n_uniform}
depends on position only through $\Dlt(r)$, which varies on the scale
$\xic$, it only contributes smooth variation in charge density.  All rapid variation of the
density must therefore come from the bound states.
The bound-state share is given by Eqs.~\eqref{eq:main}
and~\eqref{eq:outer}.  It is positive definite at every radius, being a
sum of squared hole amplitudes with positive weights $C_{\mu}^{2}>0$.

The remaining piece of $\delta n_{\rm total}$ is the response of the
density in Eq.~\ref{eq:n_uniform} to the potential they generate.  Following
Ref.~\cite{PhysRevLett.77.566} this is $(\partial n/\partial\EF)e\phi$,
which is Thomas--Fermi screening: $\kTF^{2}=4\pi
e^{2}(\partial n/\partial\mu)/\varepsilon$.  Moving it to the left-hand
side of Eq.~\eqref{eq:poisson_bare} gives the screened form
\begin{equation}
  \bigl(\nabla^{2}-\kTF^{2}\bigr)\phi(r)
  =-\frac{4\pi e}{\varepsilon}\,\delta n_{\rm bound}(r),
  \label{eq:poisson}
\end{equation}
with $\delta n_{\rm total}=\delta n_{\rm bound}
-(\varepsilon\kTF^{2}/4\pi e)\phi$.  The two equations are one, written
either with the measured charge as source or with the bound-state part
alone.




\subsection{Screened response}

The operator on the left of Eq.~\eqref{eq:poisson} is a low-pass
filter.  In momentum space its
inverse is $1/(q^{2}+\kTF^{2})$: a response of strength $1/\kTF^{2}$ at
$q\to0$, falling off beyond the cutoff $q\sim\kTF$.  Consequently
$\phi$ reproduces the source profile itself, divided by $\kTF^{2}$,
wherever the source varies slowly on the scale $1/\kTF$, and smooths
any feature sharper than that.  This is the entire content of the
screening problem, and it is what makes the answer elementary.

Inside the core the source is, from Eq.~\eqref{eq:main},
$\delta n_{\rm bound}=\langle C^{2}\rangle[1-J_{0}^{2}(\kF r)]$, and for
$\kF r\gg1$ the Bessel asymptotics separate it into two pieces of
sharply different character,
\begin{equation}
  \delta n_{\rm bound}\;\simeq\;\langle C^{2}\rangle
  \Bigl[\underbrace{1-\tfrac{1}{\pi \kF r}}_{\text{smooth}}
        \;-\;\underbrace{\tfrac{\sin 2\kF r}{\pi \kF r}}_{2\kF\ \text{ripple}}\Bigr].
  \label{eq:split}
\end{equation}
The smooth part varies on the scale $r$; the ripple oscillates at
$q=2\kF$.  Each drives its own response, and the responses inherit the
$r$-dependence of the terms that drive them.  Solving
Eq.~\eqref{eq:poisson} for each in turn---so that $\nabla^{2}$ acting
on the smooth response may be dropped, while on the oscillating one
$\nabla^{2}\sin 2\kF r=-4\kF^{2}\sin 2\kF r$---gives directly
\begin{equation}
  \phi(r)\;\simeq\;\frac{4\pi e\langle C^{2}\rangle}{\varepsilon}
  \left\{\frac{1}{\kTF^{2}}
   -\frac{1}{\pi \kF r}\left[\frac{1}{\kTF^{2}}
   +\frac{\sin 2\kF r}{4\kF^{2}+\kTF^{2}}\right]\right\},
  \label{eq:phi_two_term}
\end{equation} for $\tfrac{1}{\kF}\ll r\lesssim\xic$.
Each of the three terms of Eq.~\eqref{eq:split} is simply divided by
$q^{2}+\kTF^{2}$ at its own wavevector---$\kTF^{2}$ for the first two,
$4\kF^{2}+\kTF^{2}$ for the ripple---so every response carries the sign
of the source that produced it.  The constant is not a spectator: it is
the response to the uniform part of the bound charge, and it is what
removes that charge from the screened total below.
The replacement $\nabla^{2}\to-4\kF^{2}$ is not exact, since
$\sin(2\kF r)/r$ is not an eigenfunction of the radial Laplacian.  The
leading correction is in quadrature with the term it corrects, so it
shifts the phase at order $(\kF r)^{-1}$ but the amplitude only at
order $(\kF r)^{-2}$~\cite{SM}.

The lower limit on $r$ is enforced twice over.  The separation
\eqref{eq:split} is itself asymptotic: as $r\to0$ the bound-state
density vanishes as $\tfrac12(\kF r)^{2}$, while the $1/\pi\kF r$ terms
that represent it diverge.  Independently, dropping $\nabla^{2}$ from
the smooth response requires the source to vary slowly on the screening
length, which fails where $\delta n_{\rm bound}$ turns over on the
scale $\kF^{-1}$.  Because $\kTF$ and $\kF$ are comparable in a metal
the two conditions bite at the same radius~\cite{SM}.


\begin{figure*}[t]
\centering
\includegraphics[width=\textwidth]{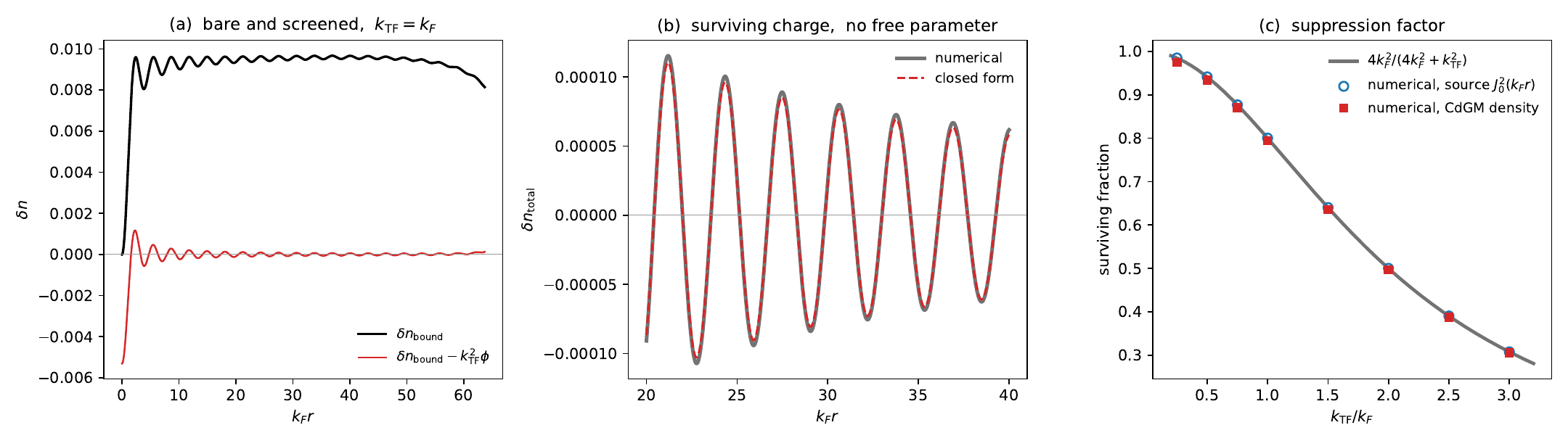}
\caption{%
Screening of the CdGM charge, computed with $\kF=1$, $\Dlt/\EF=0.01$
($\kF\xic\approx64$), using the \emph{numerical} mode sum
$2\sum_\mu C_{\mu}^{2}|f_{\mu,-}|^{2}$ as the source rather than the
closed forms.
\textbf{(a)}~The bare bound-state density (black) is positive definite
and carries a net charge; the screened total
$\delta n_{\rm bound}-\kTF^{2}\phi$ (red) carries none and alternates
in sign.  Screening removes the net charge and leaves the ripple.
\textbf{(b)}~The surviving charge against the closed form of
Eq.~\eqref{eq:screened_charge}, with no adjustable parameter.
\textbf{(c)}~The surviving fraction versus $\kTF/\kF$: curve, the
predicted $4\kF^{2}/(4\kF^{2}+\kTF^{2})$; open circles, numerical
solution with the analytic source $J_{0}^{2}(\kF r)$; filled squares,
numerical solution with the full numerical CdGM density.  The two
symbol sets test the two approximations separately---the closed-form
density and the two-term potential---and agree with the prediction to
$0.04\%$ and $1\%$ respectively.  The latter offset is the
mode-average $\langle C^{2}\rangle$ standing in for a $\mu$-dependent
$C_{\mu}^{2}$.}
\label{fig:screening}
\end{figure*}


\subsection{The surviving charge}
\label{sec:surviving}

The screened charge now follows without further work.  Inserting
Eq.~\eqref{eq:phi_two_term} into
$\delta n_{\rm total}=\delta n_{\rm bound}-(\varepsilon\kTF^{2}/4\pi e)\phi$,
the smooth parts cancel identically
while the ripple survives with a reduced weight,
\begin{equation}
  \delta n_{\rm total}(r)\;\simeq\;-\,\langle C^{2}\rangle\,
  \frac{4\kF^{2}}{4\kF^{2}+\kTF^{2}}\;
  \frac{\sin 2\kF r}{\pi \kF r}
  \label{eq:screened_charge}
\end{equation}
in the window $1/\kF\ll r\lesssim\xic$.  The physical content is a
single statement: Thomas--Fermi screening neutralises charge completely
at long wavelength but is ineffective at the largest momentum transfer
the Fermi surface allows, so the $2\kF$ component of the bound-state
charge is the only part that is not compensated.  The factor
$4\kF^{2}/(4\kF^{2}+\kTF^{2})$ is the fraction that escapes, and it is
bounded below for any metal.  In the free-electron estimate
$\kTF/\kF=0.815\sqrt{r_s/a_{\rm B}}$, and the density parameter
$r_s/a_{\rm B}$ lies between roughly two and six across the metallic
range~\cite{ashcroft1976solid}, so $\kTF/\kF$ is confined to about
$1.2$--$2.0$ and the surviving fraction to between three-quarters and
one-half.  The $2\kF$ modulation can therefore be neither screened away
nor enhanced: it is a fixed fraction of order unity of the bare
bound-state ripple, whatever the material.  This insensitivity is a
direct consequence of $\kTF$ and $\kF$ being set by the same electron
density.

Figure~\ref{fig:screening} tests this against direct numerical solution
of Eq.~\eqref{eq:poisson}.  Two independent approximations are involved
and are checked separately.  Using the closed-form source
$J_{0}^{2}(\kF r)$ isolates the two-term potential
Eq.~\eqref{eq:phi_two_term}, which reproduces the predicted suppression
to $0.04\%$ over $0.25\leq\kTF/\kF\leq3$.  Using instead the full
numerical mode sum---exact Bessel functions inside the core, matched
BKJT amplitudes outside, and the $\mu$-dependent $C_{\mu}^{2}$---tests
the closed-form density as well, and reproduces it to $1\%$, the
residual being the mode-average $\langle C^{2}\rangle$.  The solver
itself was verified against three problems with known closed-form
solutions and by two independent solution methods.


\section{Discussion}
\label{sec:discussion}


It is worth separating this mechanism from the one that produces
Friedel oscillations around an impurity, since the two give the same
periodicity by opposite routes.  There the source is structureless---a
point charge, flat in $q$, as in the classic treatment of an impurity
in a superconductor~\cite{Fetter1965}---and the oscillation is generated
entirely by the response, through the non-analyticity of the static Lindhard
function at $q=2\kF$.  Recovering the real-space decay in that case is
delicate: it rests on the behaviour of the transform near the
singularity, and is obtained by Lighthill's theorem~\cite{lighthill1958introduction}, 
with the decay exponent fixed by
the order of the non-analyticity~\cite{gabovich1978screening,PhysRevLett.18.546,Lu_2016}.
Here the roles are exchanged.  The response is Thomas--Fermi and
carries no structure whatever, while the source itself is peaked at
$2\kF$.  Both routes lead to the same wavevector for the same reason,
that $2\kF$ is the largest momentum the Fermi surface can supply, but
only in the present case is the oscillation already present in the
charge that is being screened.  That is what allows the elementary
treatment above: no property of the dielectric function beyond its
value at two wavevectors is required.

The two routes leave different fingerprints in the amplitude.  When
the oscillation comes from the response, it originates in the
derivative of $1/\varepsilon(q)$ at the singular point and the
amplitude therefore carries the transfer function \emph{squared};
Stern's result for a point charge in two dimensions has precisely this
form, $\bigl[2\kF/(2\kF+\kTF)\bigr]^{2}$ in the present
notation~\cite{PhysRevLett.18.546}.  Here the oscillation is filtered
once rather than differentiated, and the transfer function appears to
the first power.


The quantity of focus here is the density---a sum of squared amplitudes over
occupied states---rather than a spectral function or a Green's
function.  
Density is one of the primary variables of
density-functional theory, including its extension to the
superconducting state~\cite{OGK1988}, and closed-form results for it are
scarce even in one dimension, as the calculations cited in
Sec.~\ref{sec:intro} illustrate.
Equations~\eqref{eq:main} and~\eqref{eq:outer} provide the
corresponding statement for a vortex core, a geometry in which the
density varies on the scale $\kF^{-1}$ and local-density
approximations are therefore least reliable.  Self-consistent numerical
treatments of this geometry exist~\cite{GygiSchlueter1991}, but without
an analytic density there has been nothing to check them against.  Whether a functional can
reproduce a density that vanishes at the origin for topological reasons
and oscillates at $2\kF$ thereafter seems a natural test, and one that
does not require solving the pairing problem self-consistently.

Equation~\eqref{eq:screened_charge} is the analytical counterpart of
the sign change reported numerically by Machida and
Koyama~\cite{PhysRevLett.90.077003}.  Their Fig.~1(b) contrasts
$\rho(r)-\rho_\infty$ computed with and without the Poisson equation
and finds sign alternation only in the charged case, i.e. $e\neq0$, with the
oscillation period $\pi/\kF$ and an amplitude that decreases with
increasing $\kF\xic$.  All three features follow from
Eq.~\eqref{eq:screened_charge}: the period is set by $2\kF$; the
alternation follows from neutrality; and since
$\langle C^{2}\rangle\propto(\kF\xic)^{-1}$, the amplitude scales as
$(\kF\xic\,\kF r)^{-1}$, which we have verified numerically between
$\kF\xic=10$ and $64$.  Their remark that ``simple screening of the
Thomas--Fermi type does not work in the vortex state'' can be sharpened:
Thomas--Fermi screening does apply, but only its full $q$-dependence;
using the $q\to0$ limit alone would neutralise the ripple along with
everything else.

The analysis is meant to complement rather than replace self-consistent
BdG calculations such as those of Machida and
Koyama~\cite{PhysRevLett.90.077003}.  Self-consistency reshapes the gap
profile and shifts quantitative details, but the mechanism identified
here rests only on a sharp Fermi surface, the exclusion of the $\nu=0$
channel, and $\kF\xic\gg1$, and is therefore robust against smooth
deformations of $\Dlt(r)$.  Equations~\eqref{eq:main},
\eqref{eq:outer} and~\eqref{eq:screened_charge} accordingly provide an
analytical benchmark against which self-consistent numerical densities
can be checked in the weak-coupling limit.


\section{Summary}
\label{sec:summary}

Under two approximations valid to leading order in $\Dlt/\EF$---a
mode-independent normalisation constant and $k_-\approx\kF$---the
Bessel completeness identity applied to the CdGM hole basis yields the
inner density profile, Eq.~\eqref{eq:main}, while a WKB treatment of
the BKJT amplitude outside the core gives the outer envelope,
Eq.~\eqref{eq:outer}.  The $\nu=0$ Bessel
channel is excluded by the vortex winding number, which fixes $n(0)=0$
exactly within the approximation and imprints $2\kF$ oscillations on
the core density.

The bound-state charge does not integrate to zero, so overall
neutrality requires the extended states to compensate it exactly.  That compensation is complete at long wavelength but
ineffective at $q=2\kF$, and the residue is the screened charge of
Eq.~\eqref{eq:screened_charge}: a sign-alternating $2\kF$ oscillation of
amplitude $4\kF^{2}/(4\kF^{2}+\kTF^{2})$ relative to the bare ripple.
Direct numerical solution of the Poisson equation with the full CdGM
mode sum as source confirms this to better than $1\%$.  Natural extensions include
$d$-wave pairing, finite temperature, and multi-band Fermi surfaces.



\bibliographystyle{unsrt}
\bibliography{vortex_ref}

\begin{thebibliography}{10}
\bibitem{note_coreedge}
The replacement is not optional. The slow-mode coordinate $x(r)$ is imaginary
  for $r<r_t$, so the slow-mode ansatz is undefined throughout the classically
  forbidden region, which occupies a fraction $\mu/N$ of the core radius---one
  half on average over the ladder, and $0.99$ for the highest mode. That region
  is not negligible: it supplies $2$--$11\%$ of $I_{\rm in}$ for typical modes
  and up to $76\%$ for $\mu/N\to1$. Moreover the BKJT carrier contains
  $Y_{\mu}(\krho r)$, which diverges as $r^{-\mu}$ at the origin, whereas the
  physical hole amplitude must vanish as $r^{\mu+1/2}$---the behaviour that
  enforces $n(0)=0$ and hence makes the closed-form inner density available at
  all. The price is that the inner and outer regions are represented by
  different functions, which therefore need not join continuously at $r=\xic$;
  used consistently on both sides, the BKJT ansatz is continuous there by
  construction. Numerically the resulting step in $\delta n$ is $2.6\%$ at
  $\kF\xic=64$ and $9\%$ at $\kF\xic=10$, confined to a shell of a few percent
  of $\xic$ about the core edge, and no quantitative result of this work is
  drawn from that shell. We note also that the single constant $C$ in the inner
  Bessel solutions is a choice: inside the core $\Dlt=0$ decouples the two
  components, so their amplitudes could be taken independent, and doing so
  restores continuity of $f_-$ exactly. We retain the common constant because
  the near mode-independence of $C^{2}$ is what makes the closed forms of this
  work possible, and relaxing it makes $C^{2}(\mu)$ more strongly
  $\mu$-dependent. More fundamentally, continuity of both spinor components and
  their derivatives is four conditions, and these are met automatically only at
  the exact eigenvalue, where regularity at the origin and decay at infinity
  intersect. Taking the CdGM spectrum as given therefore over-determines the
  matching, and some residual is unavoidable within any outer solution carrying
  two free constants.
\bibitem{note_turningpoint}
The exception is the immediate neighbourhood of $r_t$, where the two BKJT
  components visibly separate: the hole component dips before turning up while
  the electron component rises monotonically. The mixing angle is pinned to
  $\eta(r_t)=0$ because $Y_{\mu}(\krho r)$ is the exponentially growing
  solution inside the turning point, so any admixture would destroy regularity
  at the origin. It then switches on as $\eta\simeq(\Lambda+b^{-1})x$ with
  $x\simeq(\Dlt_{\infty}\krho/E_{\rho})\sqrt{2r_t}\sqrt{r-r_t}$, giving
  $f_{\pm}(r)\simeq J_{\mu}(\krho
  r)\mp\frac{1}{2}(\Lambda+b^{-1})(\Dlt_{\infty}\krho\sqrt{2r_t}/E_{\rho})\sqrt{r-r_t}\,Y_{\mu}(\krho
  r_t)$, a correction of infinite slope at $r_t$. Since
  $Y_{\mu}(\mu)\simeq-0.775\,\mu^{-1/3}<0$ while
  $J_{\mu}(\mu)\simeq0.447\,\mu^{-1/3}>0$, it pulls $f_-$ down and pushes $f_+$
  up. For $\mu=5.5$: $J_{\mu}(\mu)=0.253$, $Y_{\mu}(\mu)=-0.439$, and $f_-$
  dips to $0.225$ near $r-r_t\simeq0.15$ before recovering. The exact solutions
  show no such feature; this is the BKJT ansatz failing at the centrifugal
  turning point, and is why the inner carrier is replaced by exact Bessel
  functions.
\end{thebibliography}

\clearpage
\onecolumngrid
\begin{center}
  \large\bfseries Supplemental Material
\end{center}
\twocolumngrid
\setcounter{section}{0}
\setcounter{equation}{0}
\setcounter{figure}{0}
\renewcommand{\thesection}{S\arabic{section}}
\renewcommand{\theequation}{S\arabic{equation}}
\renewcommand{\thefigure}{S\arabic{figure}}

\section{Validity of the replacement trick in solving screened Poisson equation}
\label{sec:crit}

The screened Poisson equation of the main text,
\begin{equation}
  \bigl(\nabla^{2}-\kTF^{2}\bigr)\phi
  =-\frac{4\pi e}{\varepsilon}\,\delta n_{\rm bound},
  \label{eq:sm_poisson}
\end{equation}
is linear, so a source written as a sum is answered by the sum of the
separate responses.  
For $\kF r\gg1$ the bound-state density separates into a smooth part and a
part oscillating at $q=2\kF$,
\begin{equation}
  \delta n_{\rm bound}\simeq\langle C^{2}\rangle
  \Bigl[1-\tfrac{1}{\pi\kF r}-\tfrac{\sin 2\kF r}{\pi\kF r}\Bigr].
  \label{eq:sm_split}
\end{equation}

What makes each piece elementary is a single step: where the source
oscillates at wavevector $q$, the corresponding potential is obtained from replacing the 
differential operator $\nabla^{2}$ by $-q^{2}$. As such, the
differential equation collapses to an algebraic one,
\begin{equation}
  \bigl(q^{2}+\kTF^{2}\bigr)\phi_{q}
  =\frac{4\pi e}{\varepsilon}\,\delta n_{q}.
  \label{eq:sm_algebra}
\end{equation}
For the smooth part $q\simeq0$, so $-q^{2}$ vanishes,
$\nabla^{2}$ is simply dropped, and the potential is then the source divided
by $\kTF^{2}$.  
Adding the two gives the following approximate solution to the screened Poisson equation,
\begin{equation}
  \phi(r)\simeq\frac{4\pi e\langle C^{2}\rangle}{\varepsilon}
  \left\{\frac{1}{\kTF^{2}}-\frac{1}{\pi\kF r}
  \left[\frac{1}{\kTF^{2}}+\frac{\sin 2\kF r}{4\kF^{2}+\kTF^{2}}\right]
  \right\}.
  \label{eq:sm_twoterm}
\end{equation}


Obviously, the replacement trick is not exact. The oscillating potential inherits the
form $\sin(qr)/r$ from its source, and $\sin(qr)/r$ is not an
eigenfunction of the radial Laplacian. To see the validity of the replacement trick, 
we differentiate $e^{iqr}/r$ and get,
\begin{equation}
  (\nabla^{2}-\kTF^{2})\!\left(\frac{e^{iqr}}{r}\right)
  =-\Bigl(q^{2}+\kTF^{2}+\frac{iq}{r}-\frac{1}{r^{2}}\Bigr)
    \left(\frac{e^{iqr}}{r}\right).
   \label{eq:sm_lap}
\end{equation}
The trick keeps the constant $q^{2}+\kTF^{2}$ in the parentheses and
discards $iq/r-1/r^{2}$, so one criterion covers both cases,
\begin{equation}
  \left|\frac{iq}{r}-\frac{1}{r^{2}}\right|\;\ll\;q^{2}+\kTF^{2}.
  \label{eq:sm_crit}
\end{equation}
(i) For the ripple, $q=2\kF$ dominates the right-hand side and $q/r$
the left, so the condition is $qr\gg1$.
(ii) For the smooth part, $q=0$
annihilates $q^{2}$ and $iq/r$ together, leaving $1/r^{2}\ll\kTF^{2}$,
that is $\kTF r\gg1$.  


\section{Validation for numerical solver}
\label{sec:fails}


The mode sum is evaluated with the exact inner Bessel solutions and the
Taylor-linearised BKJT outer solution, with $C_{\mu}^{2}$ obtained by
quadrature to $30\xic$.  The screened Poisson equation is solved on a
uniform radial grid by a second-order finite-difference scheme with
$\phi'(0)=0$ and $\phi\to0$ at the outer boundary.

In order to validate the numerical solver for screened Poisson equation, we shall
consider a simple example for which the closed-form solution is available. Moreover, we
can also check if the numerical results agree with the solutions obtained from the replacement trick.
Assume the source of charge is a uniform disc, $S=S_{0}\Theta(a-r)$.  A constant particular
solution together with $I_{0}$ inside and $K_{0}$ outside, matched in
value and slope at $r=a$, gives
\begin{align}
  \phi_{\rm in}&=\frac{S_{0}}{\kTF^{2}}
    -\frac{S_{0}aK_{1}(\kTF a)}{\kTF}\,I_{0}(\kTF r),\notag\\
  \phi_{\rm out}&=\frac{S_{0}aI_{1}(\kTF a)}{\kTF}\,K_{0}(\kTF r).
  \label{eq:sm_disc}
\end{align}
On the other hand, we can apply the replacement trick: treating the step {\it instead} as
smooth and setting $\nabla^{2}\to0$ gives at once
\begin{equation}
  \phi=\frac{S_{0}}{\kTF^{2}}\,\Theta(a-r)\:,
  \label{eq:sm_disc_trick}
\end{equation}
namely, the potential simply copies the source.

\begin{figure*}[t]
\centering
\includegraphics[width=0.86\textwidth]{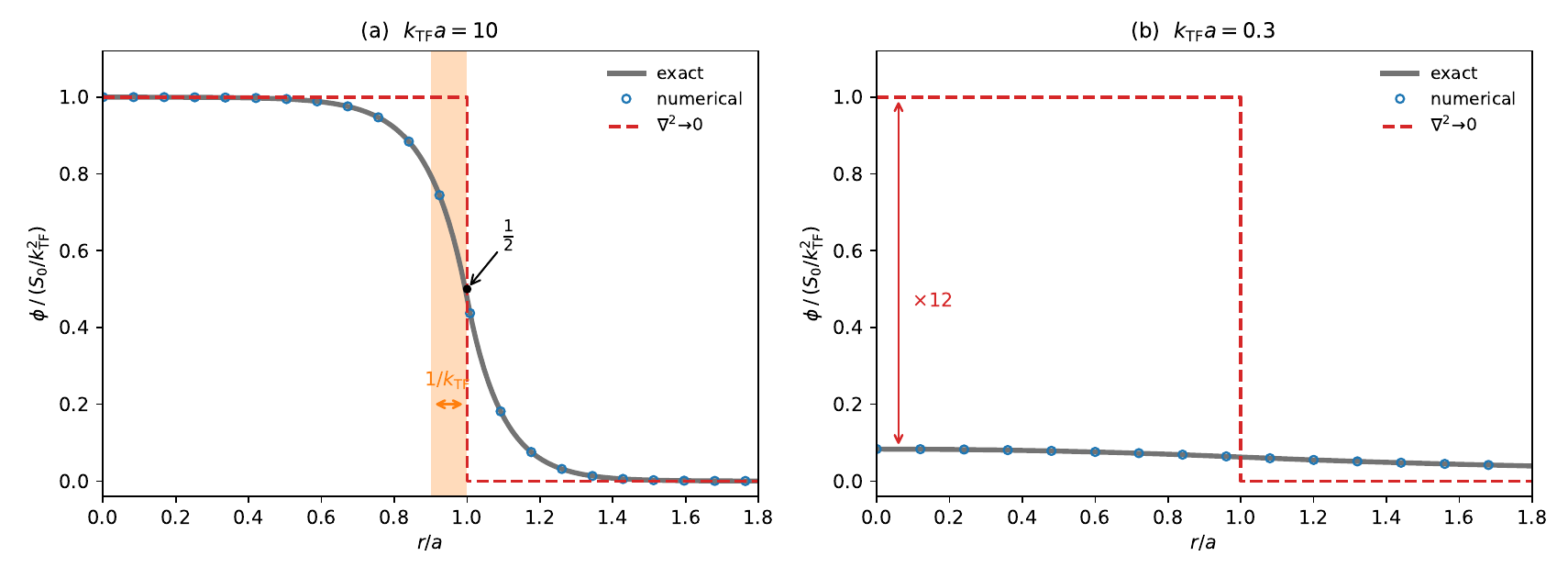}
\caption{%
The smooth replacement $\nabla^{2}\to0$ tested on a uniform disc, for
which the screened Poisson equation is solvable in closed form,
Eq.~\eqref{eq:sm_disc}.  The replacement predicts that $\phi$ copies
the source, Eq.~\eqref{eq:sm_disc_trick} (dashed).
\textbf{(a)}~$\kTF a=10$: the prediction holds through the interior and
fails only in a layer of width $1/\kTF$ at the rim (shaded), where the
potential passes through exactly half the interior value.
\textbf{(b)}~$\kTF a=0.3$: the layer is wider than the disc, and the
replacement overestimates the potential twelvefold at the origin.
Circles are the finite-difference solution, which also serves to verify
the solver against Eq.~\eqref{eq:sm_disc}.}
\label{fig:sm_disc}
\end{figure*}

Fig.~\ref{fig:sm_disc} demonstrates that the numerical solutions match well with the 
exact solution for both conditions: in panel (a) where the parameter $\kTF a=10$ and 
panel (b) where $\kTF a = 0.3$. Thus, the evidence validates the numerical solver for
screened Poisson equation.
Moreover, the solution $\phi$ obtained from the replacement
trick agrees well with numerical and analytical ones for most of the regions in panel (a) 
except for a narrow region of width of about $(1/\kTF)$.
However, the replacement-trick solution for the case in panel (b) completely fails. 
Below, we shall discuss why and where should the replacement trick fail.

Where should this fail?  The criterion \eqref{eq:sm_crit} is general,
but the condition $\kTF r\gg1$ drawn from it in
Sec.~\ref{sec:crit} was obtained by evaluating it on the source
$a\,e^{iqr}/r$.
The replacement discards $\nabla^{2}\phi$ against
$\kTF^{2}\phi$, and on the trick's own answer
Eq.~\eqref{eq:sm_disc_trick} the discarded term vanishes
\emph{identically} wherever $\phi$ is constant, and is
singular at $r=a$. 
The failure is thus confined to the rim, with no condition on $r$ at all---in
particular none at the centre.  Its width follows from the operator
alone: the homogeneous solutions of $(\nabla^{2}-\kTF^{2})\phi=0$ vary
as $e^{\pm\kTF r}$ asymptotically, so a mismatch created at $r=a$ can only heal over
$1/\kTF$, and the interior error should fall off as
$e^{-\kTF(a-r)}$.  The exact solution confirms both statements and
supplies the coefficient.  Writing it against
Eq.~\eqref{eq:sm_disc_trick},
\begin{equation}
  \frac{\phi_{\rm in}(r)}{S_{0}/\kTF^{2}}
  =1-z\,K_{1}(z)\,I_{0}(\kTF r),
  \label{eq:sm_disc_ratio}
\end{equation}
with $z\equiv\kTF a$. Equation~\eqref{eq:sm_disc_ratio} is valid inside the disc, so the error made 
by Eq.~\eqref{eq:sm_disc_trick} is
the single product $z K_{1}(z)I_{0}(\kTF r)$. Two limits settle the
question.

For $z\gg1$ the deviation is exponentially suppressed by the factor,
$zK_{1}(z)\sim z\,e^{-z}\sqrt{\pi/2z}$, and grows only on approaching
the rim, where $I_{0}(\kTF r)/I_{0}(z)\simeq e^{-\kTF(a-r)}$.  The
error is therefore confined to a boundary layer,
\begin{equation}
  1-\frac{\phi_{\rm in}}{S_{0}/\kTF^{2}}\;\simeq\;
  \tfrac12\,e^{-\kTF(a-r)},
  \label{eq:sm_layer}
\end{equation}
of width $1/\kTF$: the screening length is the distance over which the
potential can heal a discontinuity in the source.  
Everywhere deeper than a few $1/\kTF$ inside,
Eq.~\eqref{eq:sm_disc_trick} is accurate---at the centre itself to
$2\times10^{-4}$ for $z=10$, although $\kTF r$ vanishes there.  

For $z\lesssim1$ the same product behaves oppositely.  Now
$zK_{1}(z)\to1$ and $I_{0}(\kTF r)\to1$ together, so the right-hand
side of Eq.~\eqref{eq:sm_disc_ratio} approaches zero \emph{everywhere},
the origin included: the boundary layer is wider than the disc and has
nothing left to be a boundary layer of.  The replacement then fails not
by a correction but by an order of magnitude.  

\section{Error analysis of the replacement trick}

Write any radial function oscillating at wavevector $q$ as
$\phi=\mathrm{Im}\!\left[\Psi(r)e^{iqr}\right]$, separating a rapid
phase from a slowly varying complex envelope $\Psi$.  This is the same
device that underlies the BKJT ansatz for the wavefunction, where the
Hankel function $H^{(1)}_{\mu}(\krho r)$ carries the oscillation and the
spinor $g(x)$ the slow variation; here we apply it to the potential.
Using
$\nabla^{2}=d^{2}/dr^{2}+r^{-1}d/dr$ and
\begin{align}
  \bigl(\Psi e^{iqr}\bigr)' &= \bigl(\Psi'+iq\Psi\bigr)e^{iqr}, \\
  \bigl(\Psi e^{iqr}\bigr)''&= \bigl(\Psi''+2iq\Psi'-q^{2}\Psi\bigr)e^{iqr},
\end{align}
one obtains, exactly,
\begin{multline}
  \bigl(\nabla^{2}-\kTF^{2}\bigr)\phi\\
  =\mathrm{Im}\Bigl\{e^{iqr}\Bigl[\Psi''+\Bigl(2iq+\tfrac1r\Bigr)\Psi'
  -\Bigl(q^{2}+\kTF^{2}-\tfrac{iq}{r}\Bigr)\Psi\Bigr]\Bigr\}.
  \label{eq:sm_env}
\end{multline}

Equation~\eqref{eq:sm_lap} is the special case of
Eq.~\eqref{eq:sm_env} with $\Psi=1/r$ and $\kTF=0$: substituting
$\Psi'=-1/r^{2}$, $\Psi''=2/r^{3}$ gives
$\bigl[1/r^{3}-iq/r^{2}-q^{2}/r\bigr]e^{iqr}$, whose imaginary part
reproduces Eq.~\eqref{eq:sm_lap} term by term.  The advantage of
Eq.~\eqref{eq:sm_env} is that $\Psi$ is now free: instead of checking
what the operator does to a guessed function, we can solve for the
function the source demands.


Assume that the oscillating part of the source is
$\mathrm{Im}\!\left[a\,e^{iqr}/r\right]$, so
Eq.~\eqref{eq:sm_env} requires
\begin{equation}
  \Psi''+\Bigl(2iq+\tfrac1r\Bigr)\Psi'
  -\Bigl(q^{2}+\kTF^{2}-\tfrac{iq}{r}\Bigr)\Psi=-\frac{a}{r}.
\end{equation}
Trying $\Psi=A/r$ and using $\Psi'=-A/r^{2}$, $\Psi''=2A/r^{3}$, the
$r$-dependence collects into
\begin{equation}
  \frac{1}{r^{2}}-\frac{iq}{r}-\bigl(q^{2}+\kTF^{2}\bigr)=-\frac{a}{A},
\end{equation}
that is
\begin{equation}
  A(r)=\frac{a}{\bigl(q^{2}+\kTF^{2}\bigr)+\dfrac{iq}{r}-\dfrac{1}{r^{2}}}
  \;\simeq\;\frac{A_{0}}{1+i\varepsilon_{1}-\varepsilon_{2}},
  \label{eq:sm_A}
\end{equation}
with the $r$-independent $A_{0}=a/(q^{2}+\kTF^{2})$ and the
corrections $\varepsilon_{1}=q/[(q^{2}+\kTF^{2})r]$ and
$\varepsilon_{2}=1/[(q^{2}+\kTF^{2})r^{2}]$, which carry smooth spatial
variation.  The denominator of Eq.~\eqref{eq:sm_A} is the bracket of
Eq.~\eqref{eq:sm_lap} itself, and $\varepsilon_{1},\varepsilon_{2}$ are
just its two discarded terms divided by the term retained: this section
solves the same equation that Sec.~\ref{sec:crit} only inspected.
Setting $\varepsilon_{1}=\varepsilon_{2}=0$ recovers the result of the
main text.


%
The two corrections sit at different orders in $r$:
$\varepsilon_{1}\sim(r)^{-1}$ while $\varepsilon_{2}\sim(r)^{-2}$.
Expanding in powers of $1/r$ accordingly, and discarding
$\varepsilon_{2}^{2}=\mathcal{O}(r^{-4})$,
\begin{equation}
  \frac{|A|}{A_{0}}=1+\varepsilon_{2}-\tfrac12\varepsilon_{1}^{2}+\dots,
  \qquad
  \arg A=-\varepsilon_{1}+\dots
  \label{eq:sm_orders}
\end{equation}
Replacing $\nabla^{2}$ by $-q^{2}$ thus displaces the phase at leading
order but costs nothing in amplitude until second---and the suppression
factor of the main text is a statement about amplitude.

\end{document}